\documentclass[12pt,preprint]{aastex}

\newcommand{\kms}{$\;$km s$^{-1}$}
\newcommand{\msun}{$M_{\sun}$}
\newcommand{\rsun}{$R_{\sun}$}
 
\begin{document}
 
\title{S986 in M67: A Totally-Eclipsing Binary at the
Cluster Turnoff}

\author{Eric L. Sandquist}
\affil{San Diego State University,Department of Astronomy,San Diego, CA 92182}
\email{erics@mintaka.sdsu.edu}

\author{Matthew D. Shetrone}
\affil{University of Texas/McDonald Observatory,
P.O. Box 1337, Fort Davis, Texas 79734}
\email{shetrone@astro.as.utexas.edu}

\begin{abstract}
We have discovered that the star S986 in the old open cluster M67 has
detectable total eclipses of depth 0.08 mag for the primary eclipse
and 0.011 mag for the secondary eclipse (in $I$ only). We confirm the
detection of a third star in spectra contributing $(11.5\pm1.5)$\% of
the total light in $V$ band. The radial velocity of the third star
indicates that it is a cluster member, but it is unclear whether it is
physically associated with the eclipsing binary. Using
spectroscopic and photometric data, we deconvolve the photometry of
the three stars, and find that the primary star in the eclipsing
binary is significantly hotter than the turnoff. The two most likely
explanations are that the primary star is in a rapid phase of
evolution near core hydrogen exhaustion (associated with the turnoff
gap in M67's color-magnitude diagram), or that it is a blue straggler
created during a stellar collision earlier in the cluster's
history. Our detection of Li in the primary star tightly constrains
possible formation mechanisms in the blue straggler
explanation. Because S986 is often used to constrain tidal dissipation
models, this may imply that the strength of tidal effects is
underestimated.
\end{abstract}


\section{Introduction}
 
M67 is probably the most thoroughly studied old open cluster in the
Galaxy, thanks to its small distance from us. Typically quoted values
for the cluster's age ($4 \pm 0.5$ Gyr; Dinescu et al. 1995) place it
between the majority of known open clusters and the much older
globular clusters. A determination of the mass of turnoff stars in
this cluster would immediately provide us with a means of critically
testing the validity of theoretical stellar evolution tracks, and
checking models of various physical effects (such as convective
overshooting and chemical diffusion). However, because M67 seems to
have a high binary star fraction ($\sim 50$\%; Fan et al. 1996) and
there is definite evidence of strong dynamical interactions among
cluster members (e.g. van den Berg et al. 2001; Sandquist et
al. 2003), special care must be taken to ensure that the stars in such
a binary have not been influenced by their environment.

Accurate photometry of the star S986 [ID number from Sanders 1977;
also known as F111 (Fagerholm 1906), and MMJ 5624 (Montgomery,
Marschall, \& Janes 1993)] indicates that it falls at the turnoff in
the cluster's color-magnitude diagram (CMD), and proper motion studies
identify it as a high probability ($\ge 95\%$) member of the cluster
\citep{sanders,girard,zhao}. S986 was previously identified as a
single-lined spectroscopic binary by Mathieu, Latham, \& Griffin
(1990) with a period $P = 10.3386 \pm 0.0006$ d. Another interesting
aspect of this system is the fact that the primary is on a circular
orbit, indicating that tidal interactions between the stars have damped
out any initial eccentricity. Over time, circularized main sequence
binaries should be found at longer and longer periods as these
interactions have sufficient time to work in wider and wider binaries
(Mathieu \& Mazeh 1988). S986 is the longest period circularized
binary known in the cluster, and has probably required most of the
cluster's lifetime to become circularized.

As Mathieu et al. (1990) discuss, the secondary star in the binary
must be at least about 2 magnitudes fainter than the primary at 5200
\AA ~ because it is a single-lined system. This indicates that the
primary must contribute most of the light coming from the system, and
therefore must be very close to the turnoff mass. A lower limit on the
mass of the secondary is placed by the mass function for the binary:
approximately 0.51 \msun in the case of S986 (assuming a primary
mass of 1.25 \msun). Mathieu et al. stated that residuals of the
orbital solution correlated with phase, and that a weak secondary
cross correlation peak could be seen.  Later Melo, Pasquini, \& De
Medeiros (2001) definitively detected a secondary peak in the
cross-correlation function for the system, and measured a rotational
velocity $v \sin i = 5.3 \pm 0.6$ \kms. Careful examination of their
cross correlation function indicates that the secondary peak is rather
close to systematic radial velocity for M67.

As a byproduct of our project to monitor the partially-eclipsing blue
straggler S1082 (period 1.0677978 d), we made extensive observations
of the fields near the core of M67. The period of S986 has unfortunate
aliases with the period of Earth's rotation so that only one good
apparition of the primary eclipse occurs during nighttime hours for a
given observing site per month. This may explain the reason that
eclipses have not previously been detected for this system. In the
following sections, we describe our spectroscopic and photometric
observations and reduction, the modeling of the light curve,
and finally a discussion of the astrophysical meaning.

\section{Spectroscopic Observations and Reduction}

Because the spectroscopic observations of S986 provide very important
constraints on the properties of the stars involved, we discuss them
first. Our spectra were obtained at the Hobby-Eberly Telescope (HET)
with the High Resolution Spectrograph (HRS) as part of normal queue
scheduled observing for the period 2003-1. Our initial spectra were
taken with the purpose of investigating the detection of a secondary
peak in the cross-correlation function (CCF) by Melo et
al. (2001). The spectra were taken with the three arcsecond science
fiber with one sky fiber at 60,000 resolution.  Exposure times were
540 seconds and yielded signal-to-noise (S/N) of 28 per resolution
element.  Follow-up spectra were taken with the two arcsecond science
fiber with two sky fibers at 15,000 resolution.  Exposure times were
1000 seconds and yielded S/N of 100 per resolution element.  The
spectra were reduced with IRAF\footnote{IRAF (Image Reduction and
Analysis Facility) is distributed by the National Optical Astronomy
Observatories, which are operated by the Association of Universities
for Research in Astronomy, Inc., under contract with the National
Science Foundation.} ECHELLE scripts.

To determine radial velocities, we employed the IRAF task FXCOR
to cross-correlate the reduced spectra against a solar spectrum taken
with the same spectrograph. We confirmed the secondary peak in the
CCF, and deblended the two components using Gaussian profiles with
independent centroids and FWHM.  The region used for determining the
CCF was 5420 - 5720 \AA ~ for the 60,000 resolution spectra and 5960 -
6440 \AA ~ for the 15,000 resolution spectra. We find no discernible
radial velocity variation of the secondary peak even though we
observed the primary star at its maximum deviation from the cluster
mean, and so we conclude that it is not the secondary component of the
binary. Although we did examine the CCF for an even fainter third
component, we did not find any significant traces. The secondary peak
in the CCF sits at the cluster velocity (33.5 \kms; Mathieu \& Latham
1986), so that we conclude it is a cluster member: either a chance
superposition or a third star on a wide orbit around the inner
binary. We will denote the bright star in the binary as component Aa,
and the third star as component B. The velocities measured for these
two components are given in Table \ref{rvs}.
The FWHM for the primary component Aa was found to be consistent with
no detectable rotation ($<$ 10 \kms) given our resolution, while the
tertiary component B was found to have a small detectable rotational
broadening ($12 \pm 4$ \kms).  

The 60,000 resolution spectra were then shifted to the zero velocity
for component Aa and combined to produce a higher S/N master spectrum
that was used for an abundance analysis.  The techniques employed here
are nearly the same as those used in Shetrone \& Sandquist (2000):
equivalent widths were measured and abundances computed for each line.
The temperature, surface gravity, and microturbulence were varied to
minimize the slopes in the abundance vs. excitation potential and
abundance vs. reduced equivalent width plots, and also to force
ionization equilibrium.  As an additional constraint, we limit the
microturbulence to a value consistent with the formula given by
Edvardsson et al. (1993) as a function of surface gravity and
effective temperature to within their quoted errors (0.3 \kms).  One
departure from the analysis of Shetrone \& Sandquist (2000) was the
use of a flux contribution variable that was divided into Aa's EW to
compensate for the contribution of the B spectral component (although
the B component's spectral lines were removed the flux contribution
remains in our master Aa spectrum).  The flux contribution variable
was set by forcing the overall derived metallicity to be the same as
the cluster mean ($-0.05$; Shetrone \& Sandquist 2000).  Using this
technique we found the primary component has a $T_{eff} = 6400 \pm 50$
K, $\log g = 4.25 \pm 0.08$, $v_t = 2.0 \pm 0.1$ \kms and a flux
contribution of $0.885 \pm 0.015$.  Because the ``third light''
contribution to the system flux by component B affects the observed
depths of the eclipses described below, this measurement is an
important ingredient in determining system parameters.

\begin{figure}
\plotone{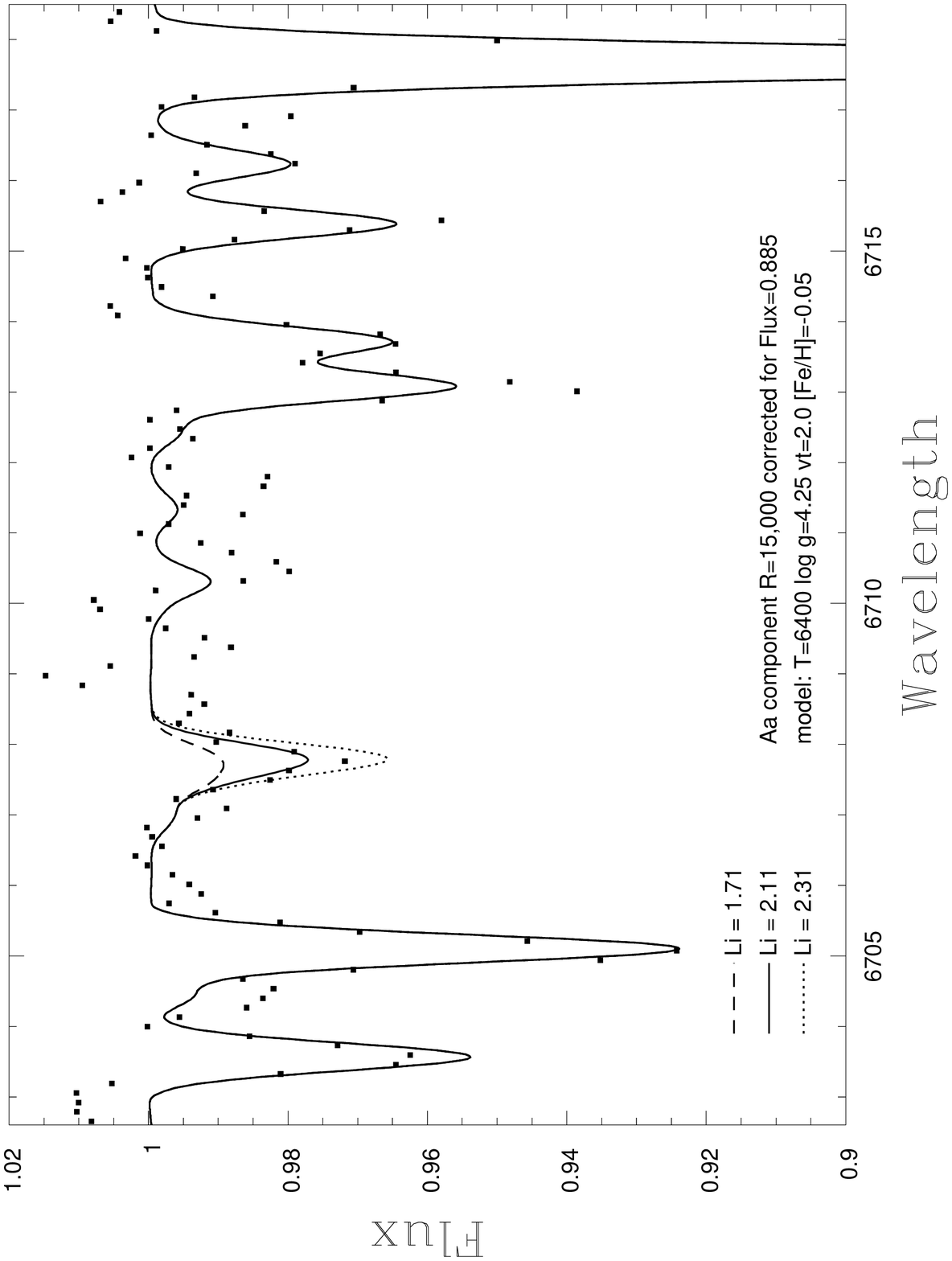}
\caption{The master combined $R = 15,000$ spectrum for component Aa in
the vicinity of the Li 6707 \AA line, with spectral syntheses for 
$\log N(\mbox{Li}) + 12 = 1.71, 2.11,$ and 2.31.
\label{li}}
\end{figure}

As part of the abundance analysis we detected a small Li line in the
master spectrum of the Aa component having an equivalent width of 10
m\AA.  This corresponds to $\log N(Li) = 2.11^{+0.2}_{-0.4}$ (see
Fig.\ref{li}), which is near the lower limit of 2.04 derived by Jones,
Fischer, \& Soderblom (1999) from data from Garc\'{\i}a Lop\'{e}z,
Rebolo, \& Beckman (1988) (although without any corrections for the
other components). In addition, our examination of O, s-process, and
$\alpha$ elements for component Aa seem to be normal for main sequence
or blue straggler stars (Shetrone \& Sandquist 2000).

A master B spectrum was created by dividing the master Aa component
into the individual spectra and then shifting the spectra to the rest
velocity of the B component and combining.  Unfortunately, because of
the low S/N of the R=60,000 spectra, individual EW could not be
measured from that master B spectrum, although the R=15,000 spectra
had sufficient signal to identify the weak B component lines (typical EW =
8 m\AA\ ).  An abundance analysis technique similiar to the one described
above was applied to this B
component but due to the small number of lines and high noise, we were
forced to constrain the surface gravity using the assumption that the
B component is a main sequence dwarf with a surface gravity and
microturbulence appropriate to its $T_{eff}$.  Using this technique we
find that the B component has $T_{eff} = 5750 \pm 200$ K with a flux
contribution of $0.097 \pm 0.025$.

\section{Photometric Observations and Reductions}

All of the photometry for this study was taken at the 1 m telescope at
the Mt. Laguna Observatory using a $2048 \times 2048$ CCD on nights
between December 2000 and November 2002. The nights of observations
are given in Table~\ref{obs}. The photometry was in $V$ and $I$ bands
with typical exposure times of 20 s (ranging between 15 and 60
s). Exposures were usually separated by about 2.5 minutes due to a
relatively long readout time for the CCD.

Most of the details of the reduction are presented in other papers
(Sandquist et al. 2003, Sandquist \& Shetrone 2003), so we only
briefly describe the reduction here.  The object frames were reduced
using overscan subtraction, bias frames, and flat fields.  We
conducted aperture photometry using the IRAF tasks DAOFIND and PHOT
from the APPHOT package.  In order to improve the accuracy of the
relative photometry for the light curves, we used an ensemble
photometry method similar to that described by Honeycutt (1992),
iterating toward a consistent solution for photometric zeropoints for
all frames and average magnitudes for all stars. Because frames
typically had more than 300 measurable stars, the formal errors in the
zero points ranged from around 0.003 to 0.007 mag, even with respect
to night-to-night variations. Our $V$ and $I$ data (given relative to
the median magnitude) are presented in Tables \ref{vdata} and
\ref{idata}, respectively.

Our photometry has been calibrated using cluster frames taken during
the same photometric night as a large number of Landolt (1992) and Stetson
(2000) standard stars. We leave the details of the calibration and the
comparison with previous datasets to a separate paper (Sandquist 2003).

\section{Light Curves}\label{analysis}

Using our observations, we first updated the ephemeris of Mathieu et
al. (1990). We have four sequences of photometric observations covering
the ingress to primary eclipse, and one set of observations covering a
portion of an egress. Because the system appears to be
non-interacting, it is reasonable to expect that the period of the
system has remained constant since the radial velocity observations
tabulated by Mathieu et al. (1990). Times of eclipse contacts were
determined by $\chi^{2}$ minimization for the light curves relative to
a trial model light curve. The times of observed first contact were
HJD 2451959.83, 2452352.68, 2452383.69, and 2452662.83, while a fourth
contact was observed at HJD 2452228.95, and one mid-eclipse point
could be determined at HJD 2452600.96. This information was used in a
simultaneous five-parameter fit to the combined dataset, where the
parameters used were period $P$, epoch of primary eclipse $T_{0}$,
system velocity $\gamma$, velocity semi-amplitude $K_{1}$, and phase
half-width of primary eclipse $\delta \phi_{p}$, assuming that the
orbit has zero eccentricity. (Tests indicated that there was no
measurable eccentricity.) The ephemeris for primary eclipse is
\[ 2445788.13(1)  + 10.33813(7) \cdot E \]
The numbers in parentheses indicate the uncertainty in the last digits
of $T_{0}$ and $P$. The period is consistent (as expected) with the
value of Mathieu et al. to within their quoted error. The velocity
parameters also match the Mathieu et al. measurements very well since
the same radial velocity data was used: $\gamma = 33.7 \pm 0.1$ \kms
and $K_{1} = 33.8 \pm 0.2$ \kms.  Finally, the fit constrains the
width of primary eclipse to be $\phi_{4} - \phi_{1} = 2 \delta
\phi_{p} = 0.0318 \pm 0.0084$. The reduced $\chi^2$ for the overall
fit was 1.32.

\begin{figure}
\plotone{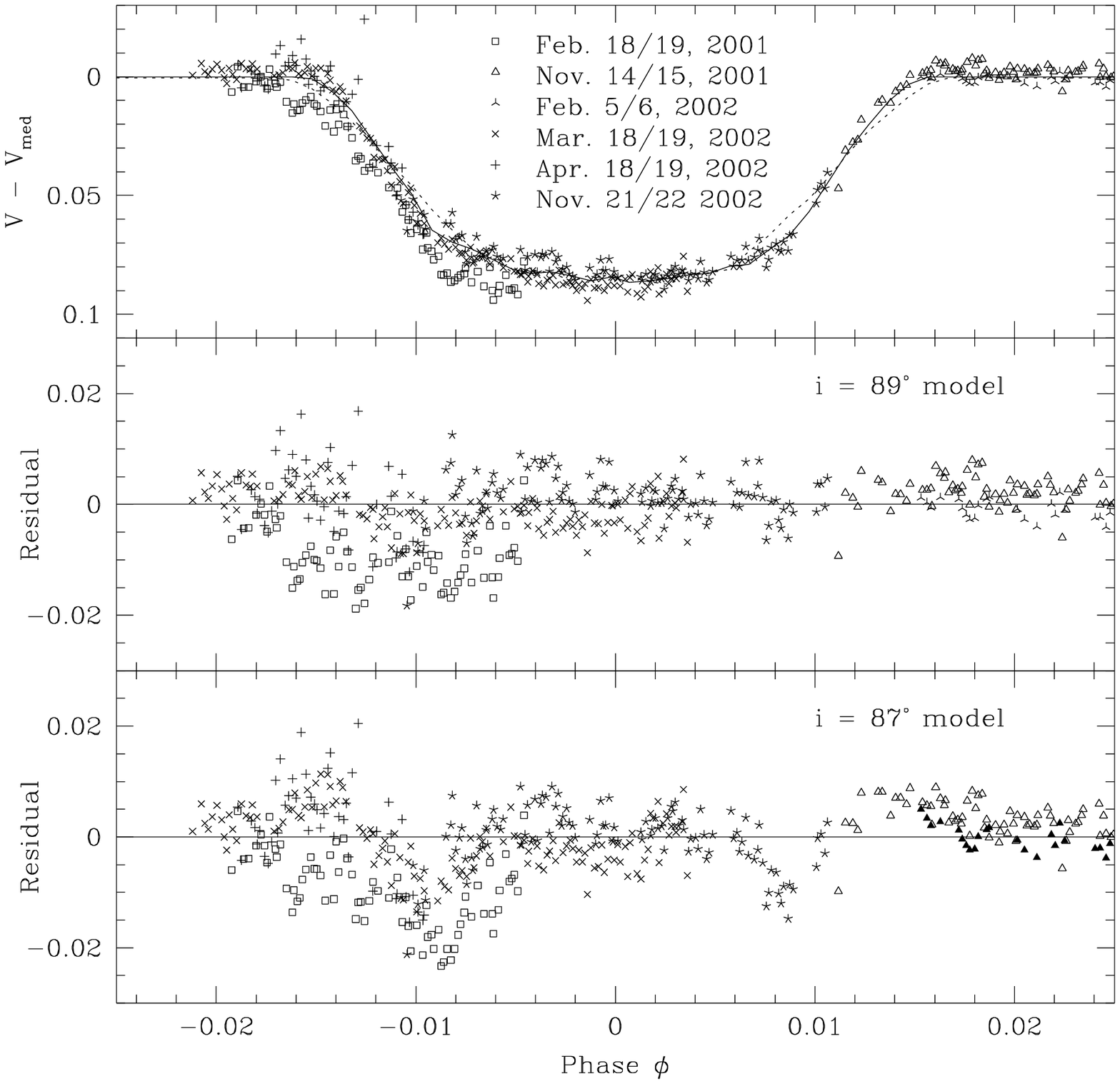}
\caption{{\it Top panel:} $V$ data for primary eclipse compared with
best fit models assuming $i = 89\degr$ {\it solid line} and
$i=87\degr$ {\it dotted line}. {\it Bottom panels:} Residuals (in the
sense of observed values minus model values) from the comparison of
the observed data with models.
\label{vprime}}
\end{figure}

\begin{figure}
\plotone{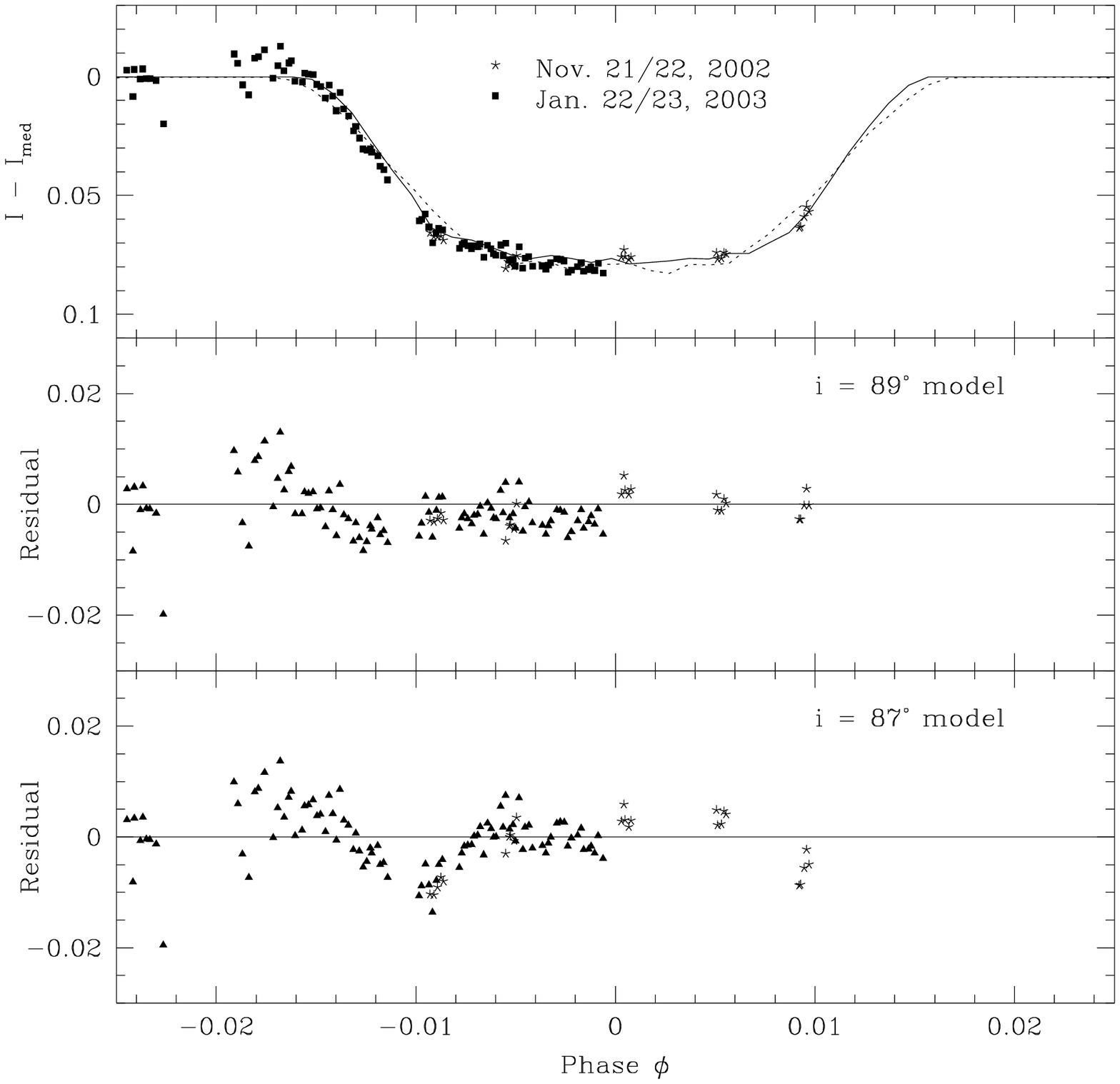}
\caption{{\it Top panel:} $I$ data for primary eclipse compared with
best fit models assuming $i = 89\degr$ {\it solid line} and
$i=87\degr$ {\it dotted line}. {\it Bottom panels}: Residuals (in the
sense of observed values minus model values) from the comparison of
the observed data with models.
\label{iprime}}
\end{figure}

\begin{figure}
\plotone{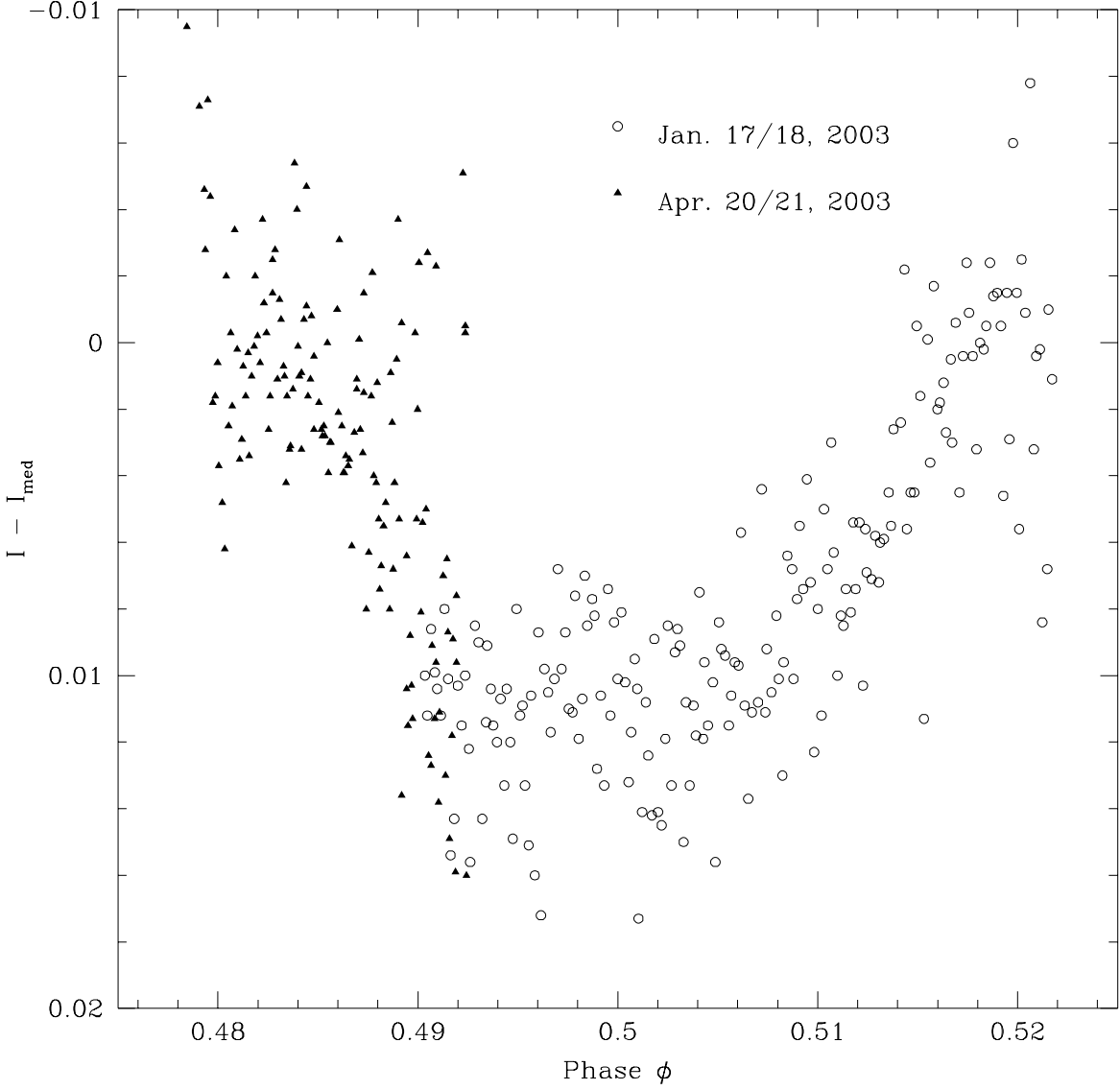}
\caption{$I$ data for secondary eclipse. The January 2003 data have
been corrected for a zero point difference of 0.0025 mag.
\label{isec}}
\end{figure}

Phased data for the phases of primary and secondary eclipse are shown
in Figs. \ref{vprime}, \ref{iprime}, and \ref{isec}. The curvature of
the light curve at second and third contacts (entering and exiting
total eclipse) indicates that the primary eclipse is indeed due to a
smaller star transiting across the face of a larger star.

\subsection{Eclipse Depths}

The eclipse depths provide robust information about the flux ratios of
the two stars (although this is affected by the third star
contributing to the system's light).  The overall depth of the primary
eclipse is approximately 0.086 mag in $V$ and 0.078 mag in $I$,
consistent with eclipse by a faint cool companion. The secondary
eclipse was not detected in $V$, but we did find evidence of a
decrease in brightness in $I$ on several nights. Data taken on the
night with best atmospheric conditions (Jan. 17/18, 2003) provided our
cleanest measurement of the depth of the secondary eclipse in $I$
($0.0111 \pm 0.0011$ mag) using the difference between the average
magnitudes out of eclipse and in total eclipse. (It appears that there
was a slight zero-point differences between the two nights, so a
single reference level in $I$ was not used. This may have come about
because the majority of our measurements in $I$ were taken during
nights of eclipse.)  We have a second measurement of the eclipse depth
($0.0101 \pm 0.0013$ mag) from Apr. 20/21, 2003. Combining these
measurements, we have a final value $0.0107 \pm 0.0008$ mag.

\subsection{Inclination and Radius Ratio}

The ratio of the radii of the two stars is also a robust measurement
because it depends on the timing of inner and outer contacts. Even
with an uncertain contribution from third light, the radii of the two
stars as determined from models are correlated, which means that the
radius ratio is minimally affected. With fairly high time resolution
observations, the times of contact can be measured directly. However,
due to the duration of the eclipses, we have not yet been able to
observe more than two eclipse contacts per night, so that some
light-curve modeling is necessary. The phase width of the primary
eclipse primarily constrains the inclination of the system and the
sizes of the two stars relative to the orbital separation. Though we
do not have a radial velocity curve for the secondary star,
the resultant uncertainties in the mass ratio $q$ do not significantly
affect the light curve fit.  Because the secondary star
acts like an almost perfectly black circular mask for most of the
purposes of the modeling the primary eclipses, uncertainties
introduced by error in the effective temperatures of the components
are also minimal.

We have used the program NIGHTFALL\footnote{See
http://www.lsw.uni-heidelberg.de/~rwichman/Nightfall.html for the
program and a user manual (Wichmann 1998)}, which allows the use of
model atmospheres (Hauschildt et al. 1999) and the modeling of
physical effects such as detailed reflection. Detailed fitting of the
primary eclipse requires some care in choosing the limb-darkening
law. We have chosen to use a two-parameter square-root law based on
literature comparisons with model atmospheres \citep{vh1,
diaz95}. Trials indicate that there are no observable indications of
ellipsoidal shapes for the components, or significant reflection
effects.

We computed light curve models covering a grid in the flux
fraction of the primary $f_{Aa} (V)$, and the inclination of the
binary $I$. We have imposed a constraint that the star contributing
the third light should be on the single star fiducial line for M67. This
assumption is not important for the determination of the inclination,
but it has a slight effect on the radius determinations, especially if
the star is itself an unresolved binary.

Although the uncertainty in the third light contribution has a
significant effect on the uncertainties in the stellar radii, the
uncertainty in the inclination dominates. The stellar radii are
correlated with the inclination, with smaller inclinations requiring
larger stellar radii to maintain the totality of the eclipse.  Based
on $\chi^{2}$ minimization, the inclination is $89\degr^{+1}_{-3}$
where the lower limit is set by when the primary eclipse unavoidably
becomes partial. A more stringent lower limit can be derived based on
when differences between observations and the best-fit model become
significant at the $3 \sigma$ level. Based on this we quote an
inclination $i = 89\degr^{+1}_{-2}$. In Figs. \ref{vprime} and
\ref{iprime}, we compare our data with best fit models of two
different inclinations.  The model fits at lower inclination first
begin to become unsatisfactory at the contact points ($\phi \approx
\pm 0.015$ and $\pm 0.01$).  Using this constraint, the ratio of the
stellar radii is $R_{Ab} / R_{Aa} = 0.268^{+0.015}_{-0.004}$, where
the uncertainties include the uncertainty in $f_{Aa} (V)$.

\subsection{Photometry of the Primary Star and its Identity}

\begin{figure}
\plotone{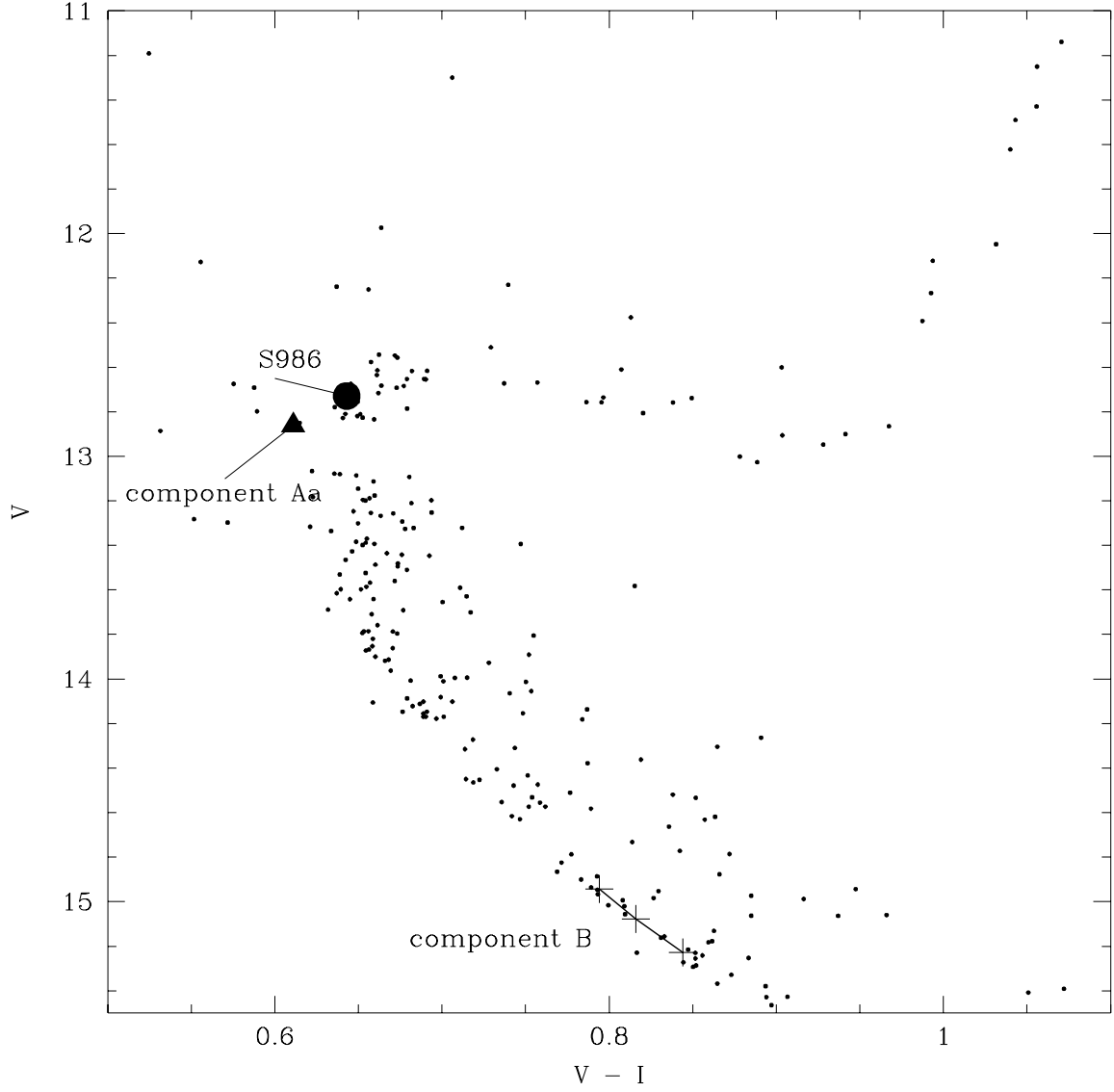}
\caption{Color-magnitude diagram for the turnoff of M67. The diagram
has been cleaned of non-members using the proper motions of
\citet{sanders}. 
\label{cmd}}
\end{figure}

The measured $T_{eff}$ and flux contribution from component B
(providing the third light) are consistent with it being a normal main
sequence star, based on comparison with \citet{yy} and \citet{gir00}
isochrones using a distance modulus $(m - M)_V = 9.72$ (Sandquist
2003). In the following analysis, we will assume that component B is a
normal main sequence star and a cluster member. Based on that, we can
constrain the photometric properties of the primary star in the
eclipsing binary (component Aa) without resorting to models. Our
calibration of the ensemble photometry (Sandquist 2003) leads to a
value for the sum of the three stars of $V_{tot} = 12.729$ and
$(V-I)_{tot} = 0.643$.  Fig. \ref{cmd} shows the $VI$ CMD for our
deconvolution of component B from the total light of the eclipsing
binary. The three linked points delineate the range of $f_{Aa} (V)$
allowed by our spectra. Our derived values are $V_{bin} = 12.86 \pm
0.02$ and $(V-I)_{bin} = 0.619 \pm 0.002$.

Assuming again that the secondary component of the eclipsing binary Ab
is a normal main sequence member of the cluster, we can deconvolve the
(small) contribution of its light and get an estimate of the
photometry of the primary alone. From the measured secondary eclipse depth
and estimates of the relative flux contributions of
components Aa and B, we find $I_{Ab} - I_{Aa} = 4.85$, or $I_{Ab} =
17.10^{+0.09}_{-0.08}$. Component Ab essentially does not contribute
in $V$ (only about 0.004 mag), so that the corrected primary star
values are $V_{Aa} = 12.86 \pm 0.02$ and $(V-I)_{Aa} = 0.611 \pm 0.003$.

The deconvolved photometry values for the primary place it
significantly to the blue of the most densely populated portions of
the turnoff. We have 2152 observations of S986 in $V$ and 816 in $I$
and comparable numbers of observations of many other turnoff stars, so
that error in the photometry relative to turnoff stars can be ruled out.
If we are wrong in our assumption that component B is a
normal main sequence star and it is actually an unresolved binary
itself, this would make B redder and would require component Aa to be
bluer still. The effective temperature measured from our spectra
indicates that it is hotter than other stars that are clearly on the
single-star sequence in the color-magnitude diagram.

\begin{figure}
\plotone{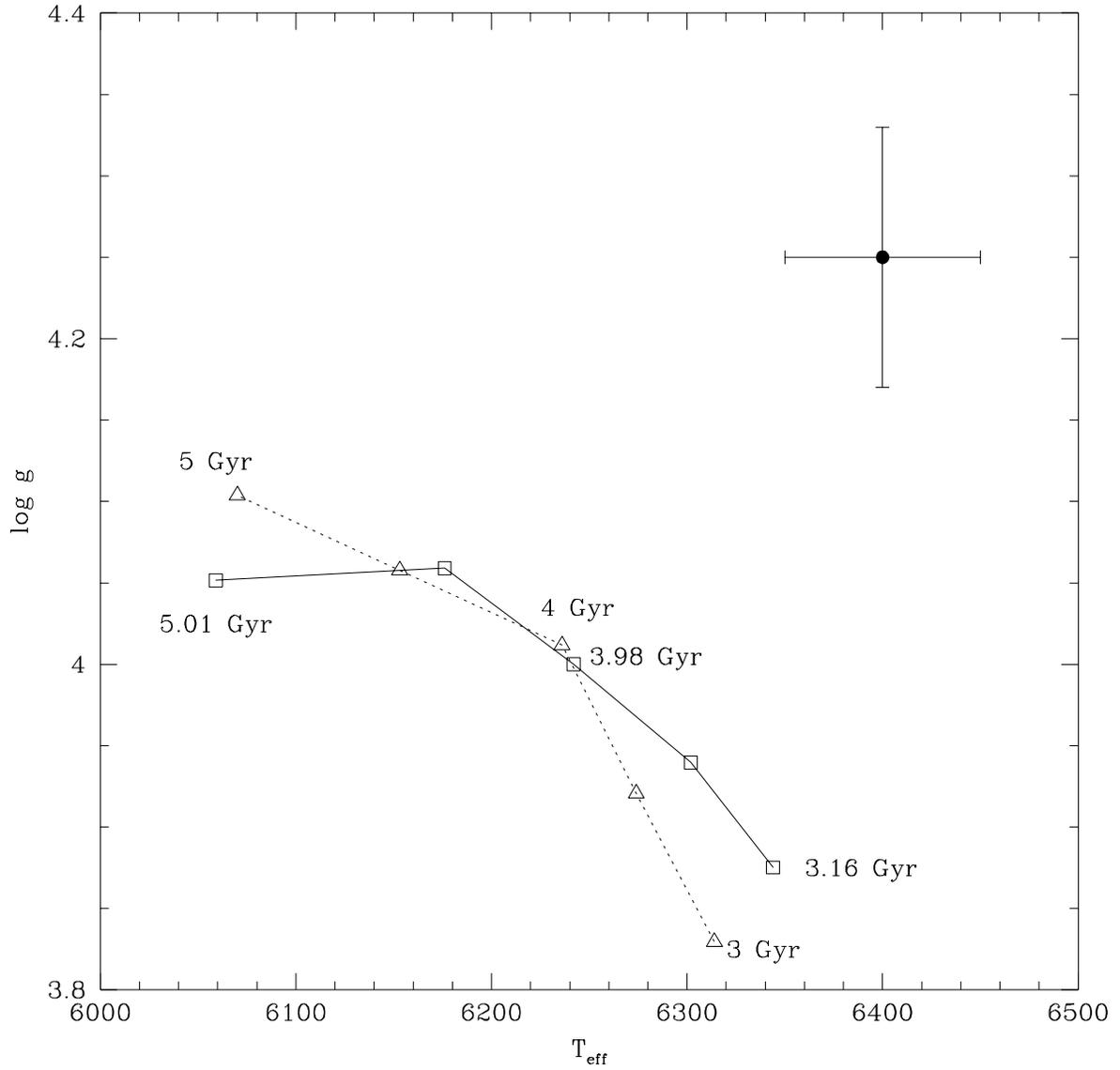}
\caption{Comparison between the observed spectroscopic properties of
component Aa of S986 and the turnoff points (maximum $T_{eff}$) from
version 2 of the theoretical isochrones of \citet{yy} ({\it dotted
line, triangles}) and \citet{gir00} ({\it solid line, squares}) with
similar amounts of convective overshooting.\label{speccomp}}
\end{figure}

We can compare our observations with theoretical models based on the
spectroscopic $T_{eff}$ and $\log g$, thereby avoiding complications
with color-effective temperature relations. In Fig. \ref{speccomp}, we
compare against the points that have the highest $T_{eff}$ on the
theoretical isochrones of Girardi et al. (2000; hereafter Padova) and
Yi et al. (2001; hereafter Y$^2$). Component Aa is clearly
inconsistent with the turnoff points from either set of isochrones,
although neither set of isochrones provides an entirely satisfactory
fit to the turnoff in the color-magnitude diagram. Although the chosen
amount of convective overshooting in the stellar models does affect
this, comparisons of the photometry with isochrones indicates that the
amount of overshooting used in the models is slightly too large for
M67 (Sandquist 2003). A smaller amount of overshooting would tend to
increase the difference between component Aa and the model turnoffs.
However, if component Aa is a blue straggler, then its surface gravity
is likely to be larger than that of a single turnoff star (which would
have evolved farther from the zero-age main sequence), consistent with
the observations.

\subsection{Secondary Star Properties\label{secprop}}

Although component Aa does not appear to be on the single-star
sequence for M67, it is quite close, and we can use that fact to
constrain the properties of the fainter component Ab. An absolute
comparison with theoretical models for low-mass stars is impossible
because the secondary has only been detected by its eclipse in $I$
band, data on the eclipsing binary does yet allow a complete solution
of the system, and the calibrated color range for our photometry does
not cover the color of component Ab. In addition, there are
substantial model differences in the treatment of surface boundary
conditions for low-mass models (Chabrier \& Baraffe 1997), and
theoretical uncertainties involving $T_{eff}$-color transformations
(Houdashelt, Bell, \& Sweigart 2000; Yi et al. 2001). In particular,
there is strong evidence that models for low mass stars become too
blue on the lower main sequence (von Hippel et al. 2002; Sandquist
2003).

\begin{figure}
\plotone{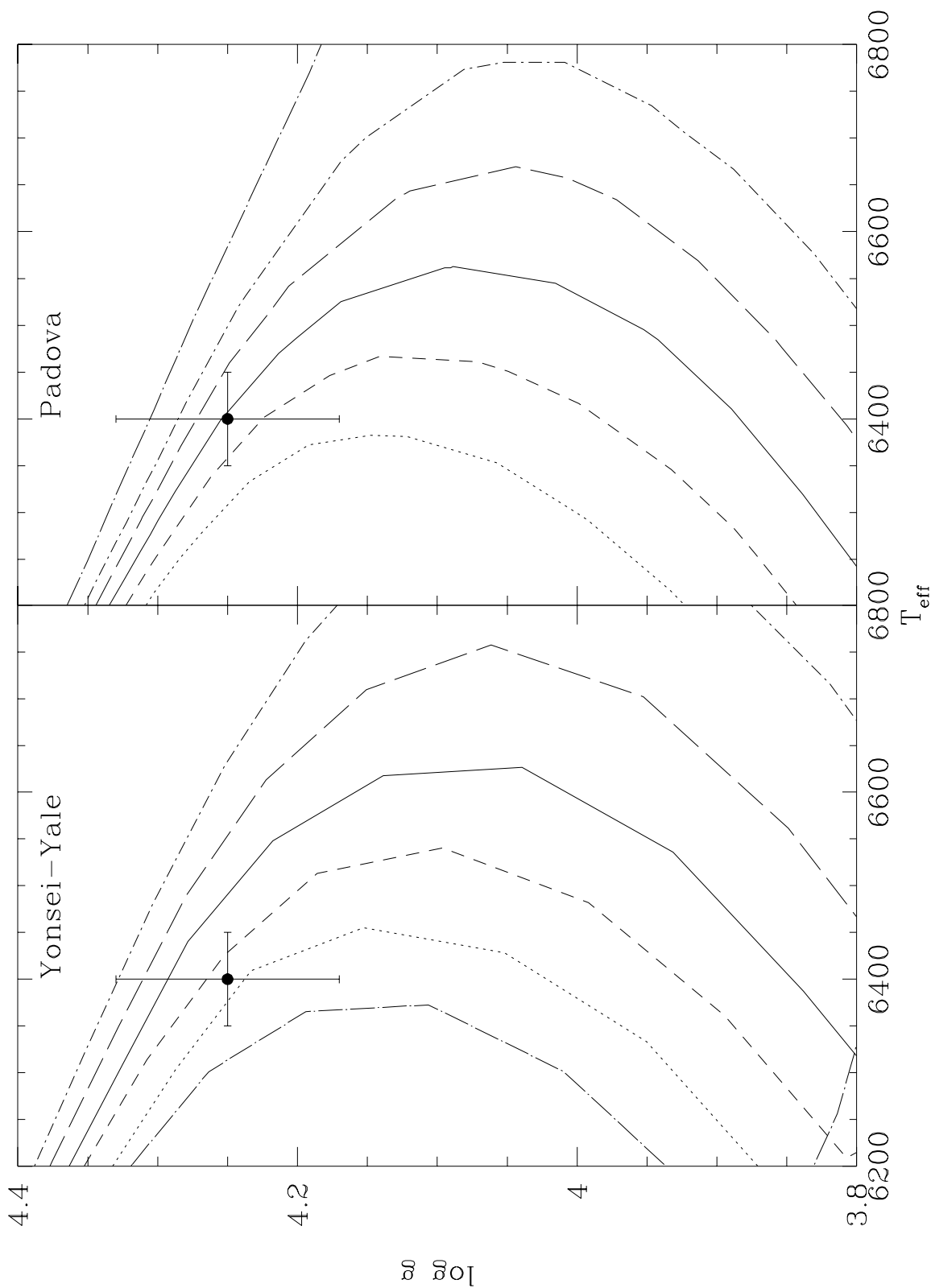}
\caption{Comparison between isochrones and spectroscopic observations
of component Aa ({\it solid circle}). From left to right, the
theoretical isochrones have ages 2.75, 2.5, 2.25, 2, 1.75, 1.5, and
1.25 Gyr for Yonsei-Yale isochrones \citep{yy}, and 2.512, 2.239,
1.995, 1.778, 1.585, and 1.259 Gyr for Padova isochrones
\citep{gir00}.\label{tgcomp2}}
\end{figure}

The most direct and robust information about component Ab comes from
the radius ratio $R_{Ab} / R_{Aa}$ and the eclipse depth in $I$ band
$\Delta I = I_{Aa} - I_{Aab}$. The value of the eclipse depth,
corrected for the third light contribution, is $\Delta I = 0.0124 \pm
0.0009$. Thus, any model comparisons we make have to be relative to
component Aa. We used isochrones of \citet{gir00} and \citet{yy} to
find models that were consistent with the spectroscopic $T_{eff}$ and
$\log g$ of component Aa in order to partially avoid uncertainties in
$T_{eff}$-color transformations. These comparisons are shown in
Fig. \ref{tgcomp2}. The properties of component Aa are consistent with those
of a star less than about 2.6 Gyr old, which can be understood if the star was
formed in a collision of lower-mass stars earlier in the cluster's history.

The isochrones allow us to get an estimate of the mass and radius of
component Aa using the spectroscopic gravity, although details of the
distribution of the star's chemicals introduce some uncertainty
(e.g. Sills et al. 1997). This is possible because of the mass-radius
relation that exists along each isochrone, with younger isochrones
giving the overall largest mass and radius values. We find nearly
identical ranges from both the \citet{yy} and \citet{gir00}
isochrones: $M_{Aa} = 1.24^{+0.19}_{-0.08}$ \msun ~ and $R_{Aa} =
1.37^{+0.24}_{-0.16}$ \rsun. We can also estimate the $I$-band
magnitude of component Aa using the isochrones and the spectroscopic
measurements, and again we find similar ranges from the two sets of
isochrones ($M_{I,Aa} = 3.08^{+0.20}_{-0.28}$). The primary source of
uncertainty in all three estimates is from the measurement of the
surface gravity.

\begin{figure}
\plotone{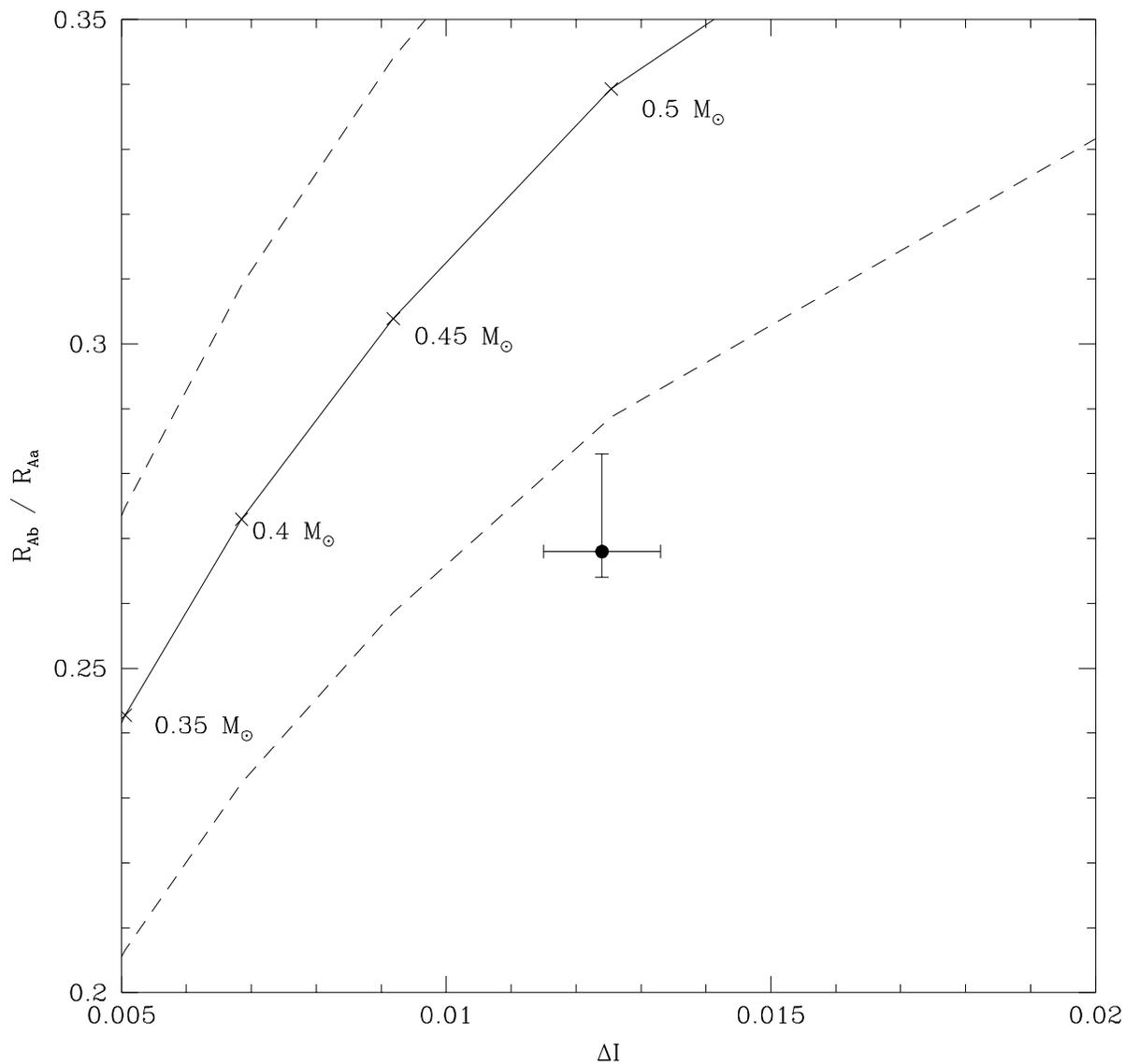}
\caption{Comparison between the radius ratio $R_{Ab} / R_{Aa}$ and
secondary eclipse depth $\Delta I$ for S986. The solid circle is the
observed value. Theoretical values derived from the low-mass stellar
models of \citet{bcah} and the isochrones of \citet{gir00} and
\citet{yy} are shown.  The {\it dashed lines} show the uncertainties
in the theoretical values due to uncertainties in the properties of
component Aa.\label{eclcomp}}
\end{figure}

In order to make the final comparison with our observed values from
the eclipse analysis, we use the models of Baraffe et al. (1998;
hereafter BCAH) since they currently provide the best fit to the lower
main sequence of M67 (although they begin to become too blue by $V
\approx 17$; Sandquist 2003).  The BCAH models are also the ones that
have been most thoroughly compared to observed colors of low-mass
stars, and seem to agree well down to $T_{eff} \sim 3600$ K (BCAH).
The isochrones of the Y$^2$ and Padova groups employ different tables
to convert from $T_{eff}$ to color, and get widely different
results. The BCAH models most nearly agree with those of the Y$^2$ or
Padova groups near solar mass, where the agreement is forced by
calibration to the Sun. We made a correction of 0.077 mag in $I$ to
account for zero-point differences between the Padova and Y$^2$
isochrones and those of BCAH for solar-mass models. Such a correction
is justified by current uncertainties in the zero-points of
$T_{eff}$-color relations for different sets of isochrones. Our final
comparison is shown in Fig. \ref{eclcomp}. The figure indicates that
component Ab is smaller and/or brighter in $I$ band than the models
predict. The uncertainties in the spectroscopic properties of
component Aa mean, however, that the inconsistency is only at a little
over $1 \sigma$.

A reasonable estimation of the effective temperature of star Ab is
next to impossible, partly because there are not enough well-measured
cool dwarf stars to set an empirical $T_{eff}$-color relation or
calibrate a theoretical one. In addition, there is some evidence that
the giant stars typically used to calibrate relations at low
temperatures may follow a systematically different relation than the
dwarfs \citep{houda}. The depth of secondary eclipse is sensitive to
$T_{eff}$, but the interpretation {\it requires} the use of updated
stellar atmosphere models because of strongly non-grey nature of the
atmospheres of low-mass stars (e.g. Chabrier \& Baraffe 1997). From
the BCAH models and a distance modulus $(m - M)_{I} = 9.67 \pm 0.03$
(Sandquist 2003), we roughly estimate that $T_{eff} = 3700 \pm 200$ K,
but we again note that this is near where their models start to
diverge from color-magnitude diagrams of field dwarfs. This
corresponds to a mass $M_{Ab} \sim 0.52$ \msun, so that $q \sim 0.41$.

\section{Discussion and Conclusions}

We have presented the first evidence of total eclipses (including
eclipses of the faint secondary star in $I$ band) in the system S986
using an extensive series of photometric observations. We have taken
high-resolution spectroscopy of the system and have identified a third
star that contributes to the light of the system. The third star
appears to be a cluster member, but may or may not be physically
associated with the eclipsing binary. The results of our analysis are given
in Table \ref{props}.

The detailed analysis of our spectroscopy and photometry for the S986
indicates that the primary star in the eclipsing binary (component Aa)
is a star that is slightly (but significantly) hotter than the turnoff
of the cluster. Two stellar explanations for this exist.  One
possibility is that the primary is a normal main sequence star that is
in a relatively short lived phase of its evolution. The gap in the CMD
of M67 with $12.85 < V < 13.1$ corresponds to a rapid phase before and
after core hydrogen exhaustion during which the central convection
zone disappears and a shell fusion source is established. The size of
the gap depends on exactly when the evolutionary timescale is small
enough that few or no stars are likely to be found in the phase, given
the total population of stars in the cluster. Several other members of M67
exist in the same portion of the CMD as component Aa (S489, S602,
S610, S615, S1271, S1503, and S1575; Sandquist 2003), which may lend
some credence to this idea. Because turnoff stars in clusters of M67's age 
have small but significant convective cores, they put difficult constraints on 
the theory of convective overshooting that have not been satisfied as yet.
Once they are, the CMD position of component Aa can be re-evaluated.

A second explanation is that component Aa is a blue
straggler.  It is somewhat difficult to piece together a scenario that
can explain the system's orbital and photometric properties, but the
light curve analysis makes it clear that component Ab is a relatively
normal main sequence star and that the two stars are completely
detached. Component B is also consistent with being a normal main
sequence star. As such, scenarios involving any kind of mass transfer
to star Aa are unlikely.

Our detection of Li in component Aa is consistent with the abundances
of stars on the edge of the ``Li gap'' (centered at $T_{eff} \approx
6700$ K).  On the other hand, there has not been a detection of Li in a
blue straggler in M67 to date. If component Aa was created in the
merger of two stars, the detection of Li would require that the more
massive star to have retained a substantial amount of surface lithium
in the time before the collision, that little mass was lost from the
surface of the star in the merger (see Lombardi et al. 2002 for a
discussion), and the remaining Li was not substantially depleted after
the merger. For a star with $T_{eff} = 6400$ K, the Li depletion
timescale seems to be several Gyr (Jones et al. 1999), comparable to
the age of M67, so post-merger depletion is probably not large enough
to matter here. For a star to have retained surface Li for
approximately 1.5 Gyr, $T_{eff}$ could be as low as about 5700 K based
on observations of NGC 752 (Hobbs \& Pilachowski 1986). This
corresponds to a mass of approximately 1 \msun, meaning that the other
star involved would had to have been low in mass ($\la 0.25$
\msun). A blue straggler scenario is possible with the detection of
Li, but the possible formation routes are fairly tightly constrained.

One of the most interesting details about the S986 system is that it
is circularized in spite of its relatively long period and the main
sequence nature of its component stars. The tidal circularization
places a lower limit on the age of the binary, although the limit is
not likely to be very strong given the state of our knowledge of tidal
interactions (e.g. Terquem et al. 1998) and that S986 is often used to
constrain the strength of tidal dissipation (after assuming an age for
M67). Our best blue straggler scenario for the S986 system involves a
collision of relatively low-mass stars during a close single-binary or
binary-binary interaction. During the the interaction two of the stars
could merge to form the primary star, while a third star could be
captured onto a relatively close orbit. Because strong multiple star
interactions typically produce eccentric orbits, a period of
circularization would still be necessary, which still should have
taken most of the age of the cluster. However, this would also have
the benefit of explaining why component Aa is so close to the cluster
turnoff: the merger would occur before either of the stars had done
much nuclear processing of their material. (If the input stars were
both of low mass, this would also help.) Our isochrone comparisons
imply that the merger probably happened more than 1.4 Gyr
ago. Component B may or may not be associated with the creation of the
system in this scenario. Long-term radial velocity monitoring would
help answer that question.

Whether or not component Aa is a normal main sequence star or a blue
straggler, further study of the binary is worthwhile because it
provides a means of simultaneously comparing the properties of a
low-mass main sequence star and a star with slightly more mass than
the Sun. Because the structures of these two types of stars are rather
different, theoretical models often require
assumptions that can result in systematic errors when isochrones are
computed.  In the case of an open cluster like M67, the degree of
convective core overshooting is an excellent example since it
substantially affects the evolution of stars slightly more massive
than the Sun (having relatively small but significant convective
cores), but has negligible effect on lower main sequence models, which
have surface convection zones of large extent. The surface boundary
condition is also a potential source of problems because it sets the
interior structure of fully- or mostly-convective stars while having a
minor effect on models of stars with mass near that of the Sun
(Chabrier \& Baraffe 1997). M67 provides an excellent testing ground for 
these kinds of stellar physics issues.

\acknowledgments

E.L.S. would like to thank the director of Mount Laguna Observatory
(P. Etzel) for generous allocations of telescope time to constrain the
ephemeris of this binary system, L. Detweiler for allowing us to use
one of his nights to observe an eclipse, and P. Etzel and D. Latham
for useful conversations. This research was supported in part by the
National Science Foundation under Grant No. AST-0098696 to E.L.S.

\begin{deluxetable}{lcrc}
\tablewidth{0pt}
\tablecolumns{4}
\tablecaption{Measured Radial Velocities for Components of S986}
\tablehead{\colhead{HJD} & \colhead{Resolution} & \colhead{$v_{Aa}$ (\kms)} & \colhead{$v_{B}$ (\kms)}}
\startdata
2452329.6377 & 60k & $69.6\pm0.5$ & $36.0\pm4.4$\\
2452619.0219 & 60k & $68.6\pm0.5$ & $35.2\pm4.5$\\
2452621.8263 & 60k & $31.1\pm0.5$ & \\
2452634.9772 & 60k & $1.3\pm0.5$ & $34.1\pm3.4$\\
2452659.9042 & 60k & $67.7\pm0.4$ & $39.6\pm5.8$\\
2452711.7820 & 15k & $65.7\pm0.3$ & $34.6\pm3.3$\\
2452714.7642 & 15k & $31.3\pm0.4$ & \\
2452722.7410 & 15k & $66.2\pm0.3$ & $34.2\pm3.0$\\
\enddata
\label{rvs}
\end{deluxetable}

\begin{deluxetable}{clccl}
\hspace*{-0.2in}
\tablecolumns{5}
\tablewidth{0pc}
\tablecaption{Observing Log for Photometry of S986 at Mt. Laguna}
\tablehead{\colhead{\#} & \colhead{Date} & \colhead{Filter}
& \colhead{mJD Start\tablenotemark{a}} & \colhead{$N_{obs}$}}
\startdata
1 & Dec. 5/6, 2000 & $V$ & 1884.878 & 60\\
2 & Dec. 6/7, 2000 & $I$ & 1885.871 & 38\\
3 & Dec. 7/8, 2000 & $V$ & 1886.806 & 111\\
4 & Dec. 11/12, 2000 & $V$ & 1890.817 & 57\\
5 & Dec. 12/13, 2000 & $V$ & 1891.820 & 116\\
6 & Jan. 23/24, 2001 & $V$ & 1933.675 & 206\\
7 & Jan. 25/26, 2001 & $V$ & 1935.673 & 176\\
8 & Jan. 29/30, 2001 & $V$ & 1939.676 & 44\\
9 & Jan. 30/31, 2001 & $V$ & 1940.657 & 130\\
10 & Jan. 31/Feb. 1, 2001 & $V$ & 1941.663 & 161\\
11 & Feb. 17/18, 2001 & $V$ & 1958.655 & 63\\
12 & Feb. 18/19, 2001 & $V$ & 1959.790 & 75\\
13 & Nov. 14/15, 2001 & $V$ & 2228.901 & 74\\
14 & Jan. 20/21, 2002 & $V$ & 2295.735 & 91\\
15 & Jan. 21/22, 2002 & $V$ & 2296.687 & 89\\
16 & Jan. 24/25, 2002 & $V$ & 2299.690 & 124\\
17 & Feb. 5/6, 2002 & $V$ & 2311.649 & 119\\
18 & Feb. 10/11, 2002 & $V$ & 2316.638 & 167\\
19 & Mar. 18/19, 2002 & $V$ & 2352.625 & 131\\
20 & Apr. 13/14, 2002 & $V$ & 2378.638 & 103\\
21 & Apr. 18/19, 2002 & $V$ & 2383.661 & 47\\
22 & Nov. 21/22, 2002 & $VI$ & 2600.845 & 108,25\\
23 & Jan. 17/18, 2003 & $I$ & 2657.718 & 192\\
24 & Jan. 22/23, 2003 & $I$ & 2662.702 & 110\\
25 & Feb. 17/18, 2003 & $I$ & 2688.617 & 147\\
26 & Mar. 20/21, 2003 & $I$ & 2719.635 & 138\\
27 & Apr. 20/21, 2003 & $I$ & 2750.638 & 150\\
\enddata
\label{obs}
\tablenotetext{a}{mJD = HJD - 2450000}
\end{deluxetable}

\begin{deluxetable}{lrcr}
\hspace*{-0.2in}
\tablecolumns{4}
\tablewidth{0pc}
\tablecaption{$V$-Band Photometry of S986}
\tablehead{\colhead{mJD\tablenotemark{a}} & \colhead{$V - V_{med}$} & 
\colhead{$\sigma_V$} & \colhead{Phase}}
\startdata
1884.8806 &   0.0156 &   0.0066 &   0.7346 \\
1884.9027 &   0.0057 &   0.0056 &   0.7367 \\
1884.9062 &  $-$0.0056 &   0.0043 &   0.7371 \\
1884.9095 &  $-$0.0036 &   0.0042 &   0.7374 \\
1884.9179 &  $-$0.0056 &   0.0041 &   0.7382 \\
1884.9198 &  $-$0.0044 &   0.0037 &   0.7384 \\
1884.9216 &  $-$0.0004 &   0.0038 &   0.7386 \\
1884.9231 &  $-$0.0057 &   0.0040 &   0.7387 \\
1884.9248 &   0.0040 &   0.0038 &   0.7389 \\
1884.9263 &  $-$0.0008 &   0.0036 &   0.7390 \\
1884.9281 &  $-$0.0008 &   0.0047 &   0.7392 \\
1884.9295 &  $-$0.0029 &   0.0040 &   0.7393 \\
1884.9324 &   0.0003 &   0.0039 &   0.7396 \\
1884.9339 &  $-$0.0019 &   0.0044 &   0.7398 \\
1884.9354 &  $-$0.0015 &   0.0036 &   0.7399 \\
1884.9370 &  $-$0.0014 &   0.0038 &   0.7401 \\
\enddata
\label{vdata}
\tablenotetext{a}{mJD = HJD - 2450000}
\tablecomments{The complete version of this table is in the
electronic edition of the Journal. The printed edition contains only a sample.}
\end{deluxetable}

\begin{deluxetable}{lrcr}
\hspace*{-0.2in}
\tablecolumns{4}
\tablewidth{0pc}
\tablecaption{$I$-Band Photometry of S986}
\tablehead{\colhead{mJD\tablenotemark{a}} & \colhead{$I - I_{med}$} & 
\colhead{$\sigma_I$} & \colhead{Phase}}
\startdata
1885.8712 &   0.0160 &   0.0033 &   0.8304 \\
1885.8800 &   0.0002 &   0.0033 &   0.8313 \\
1885.8831 &   0.0041 &   0.0033 &   0.8316 \\
1885.8847 &   0.0054 &   0.0033 &   0.8317 \\
1885.8863 &   0.0031 &   0.0026 &   0.8319 \\
1885.8891 &   0.0007 &   0.0033 &   0.8322 \\
1885.8912 &   0.0006 &   0.0041 &   0.8324 \\
1885.8926 &   0.0004 &   0.0041 &   0.8325 \\
1885.8939 &   0.0000 &   0.0026 &   0.8326 \\
1885.8963 &   0.0058 &   0.0033 &   0.8328 \\
1885.9119 &  $-$0.0012 &   0.0041 &   0.8344 \\
1885.9155 &  $-$0.0012 &   0.0033 &   0.8347 \\
1885.9184 &  $-$0.0030 &   0.0034 &   0.8350 \\
1885.9198 &   0.0042 &   0.0065 &   0.8351 \\
1885.9211 &  $-$0.0166 &   0.0057 &   0.8352 \\
\enddata
\label{idata}
\tablenotetext{a}{mJD = HJD - 2450000}
\tablecomments{The complete version of this table is in the
electronic edition of the Journal. The printed edition contains only a sample.}
\end{deluxetable}

\begin{deluxetable}{lc}
\tablewidth{0pt}
\tablecaption{Properties of the S986 System}
\tablehead{\colhead{Parameter} & \colhead{Value}}
\startdata
$P$ (d) & $10.33813 \pm 0.00007$\\
$i$ & $89 ^{+1}_{-2}$\\
$[L_{Aa} / L_{tot}]_{V}$ & $0.885 \pm 0.015$\\
$T_{Aa}$ (K) & $6400 \pm 50$ \\
$T_{Ab}$ (K) & $3750 \pm 200$\\
$T_{B}$ (K) & $5750 \pm 200$ \\
$V_{tot}$ & 12.729\\
$(V-I)_{tot}$ & 0.643\\
$V_{Aa}$ & $12.86 \pm 0.02$\\
$(V-I)_{Aa}$ & $0.619 \pm 0.002$\\
$V_{B}$ & $15.08 \pm 0.14$\\
$(V-I)_{B}$ & $0.816 \pm 0.025$\\
$I_{Ab}$ & $17.10 \pm 0.09$\\
$(V-I)_{Ab}$ & $\sim 2.0$\\
$\log g_{Aa}$ & $4.25 \pm 0.08$ \\
$v_{rot,Aa}$ (\kms) & $< 10$\\
$v_{rot,B}$ (\kms) & $12 \pm 4$\\
$R_{Ab} / R_{Aa}$ & $0.268^{+0.015}_{-0.004}$\\
\enddata
\label{props}
\end{deluxetable}

\end{document}